\documentclass[12pt]{JHEP3}
\usepackage{latexsym,amsfonts,amssymb,epsfig}

\newcommand{\ncc}{\newcommand}
\ncc{\grad}{\nabla}
\ncc{\tr}{\mathop{\rm tr}}
\ncc{\half}{{1\over 2}}
\ncc{\third}{{1\over 3}}
\ncc{\be}{\begin{equation}}
\ncc{\ee}{\end{equation}}
\ncc{\bea}{\begin{eqnarray}}
\ncc{\eea}{\end{eqnarray}}

\def\Z{\mathbb{Z}}
\def\C{\mathbb{C}}
\def\R{\mathbb{R}}
\def\Id{\mathbb{I}}

\ncc{\dint}[2]{\int\limits_{#1}^{#2}}
\ncc{\D}{\displaystyle}
\ncc{\PDT}[1]{\frac{\partial #1}{\partial t}}
\ncc{\tw}{\tilde{w}}
\ncc{\tg}{\tilde{g}}
\ncc{\newcaption}[1]{\centerline{\parbox{6in}{\caption{#1}}}}
\def\href#1#2{#2} 

\ncc{\al}{\alpha}
\ncc{\ga}{\gamma}
\ncc{\de}{\delta}
\ncc{\ep}{\epsilon}
\ncc{\ze}{\zeta}
\ncc{\et}{\eta}

\ncc{\Th}{\Theta}
\ncc{\ka}{\kappa}
\ncc{\la}{\lambda}
\ncc{\rh}{\rho}
\ncc{\si}{\sigma}
\ncc{\ta}{\tau}
\ncc{\up}{\upsilon}
\ncc{\ph}{\phi}
\ncc{\ch}{\chi}
\ncc{\ps}{\psi}
\ncc{\om}{\omega}
\ncc{\Ga}{\Gamma}
\ncc{\De}{\Delta}
\ncc{\La}{\Lambda}
\ncc{\Si}{\Sigma}
\ncc{\Up}{\Upsilon}
\ncc{\Ph}{\Phi}
\ncc{\Ps}{\Psi}
\ncc{\Om}{\Omega}
\ncc{\ptl}{\partial}
\ncc{\ov}{\overline}
\ncc{\gsl}{\!\not}
\ncc{\bi}[1]{\bibitem{#1}}
\ncc{\fr}[2]{\frac{#1}{#2}}
\ncc{\gm}{\mbox{$\gamma_{\mu}$}}
\ncc{\gn}{\mbox{$\gamma_{\nu}$}}
\ncc{\Le}{\mbox{$\fr{1+\gamma_5}{2}$}}
\ncc{\Ri}{\mbox{$\fr{1-\gamma_5}{2}$}}
\ncc{\GD}{\mbox{$\tilde{G}$}}
\ncc{\gf}{\mbox{$\gamma_{5}$}}
\ncc{\Ima}{\mbox{Im}}
\ncc{\Rea}{\mbox{Re}}
\ncc{\ntwo}{${\cal N}\!\!=\!2\;$}
\ncc{\none}{${\cal N}\!\!=\!1\;$}
\ncc{\nfour}{${\cal N}\!\!=\!4\;$}
\ncc{\nones}{${\cal N}\!\!=\!1^*\;$}
\ncc{\slz}{SL(2,$\mathbb{Z})\;$}

\def \bi{\bibitem}
\ncc{\rf}[1]{(\ref{#1})}

\title{\Large A Note on Domain Walls and the 
Parameter Space of \boldmath{\none} Gauge Theories }

\author{Adam Ritz \\ DAMTP, Centre for Mathematical Sciences, University
of Cambridge, \\ Wilberforce Rd., Cambridge CB3 0WA, 
United Kingdom \\ 
Email: \email{a.ritz@damtp.cam.ac.uk}}

\abstract{We study the spectrum of BPS domain walls 
within the parameter space of \none U($N$) gauge theories with adjoint
matter and a cubic superpotential. Using a low energy description obtained by 
compactifying the theory on $\R^3 \times S^1$, we examine the wall spectrum 
by combining direct calculations at special points in the parameter 
space with insight drawn from the leading order potential 
between minimal walls, i.e those interpolating between adjacent
vacua. We show that the multiplicity of composite BPS walls -- as
characterised by the CFIV index -- exhibits 
discontinuities on marginal stability curves within the parameter space
of the maximally confining branch. The structure of these marginal
stability curves for large $N$ appears tied to certain 
singularities within the matrix model description of the confining vacua.}

\keywords{BPS domain walls, \none supersymmetric gauge theories}

\preprint{DAMTP-2003-81}

\begin{document}

\section{Introduction}

Four-dimensional \none supersymmetric gauge theories are believed, and
in some cases have been shown, to exhibit many of the subtle and 
physically relevant phases seen within the standard model. In
particular, in the last decade technical advances have allowed us to
explore properties such as abelian confinement and chiral symmetry
breaking within specific \none theories \cite{sw,seiberg}. 
However, this progress has
generally been limited to particular examples, and one may hope that 
a more global perspective on the space of \none theories is
attainable. In this regard, interesting recent work by Ferrari \cite{f2} and
Cachazo, Seiberg and Witten \cite{csw1,csw2} has provided
insight into the structure of the quantum parameter space of \none
theories. This work was stimulated by the realisation of Dijkgraaf and
Vafa \cite{dv} that a matrix model structure apparently underlies the 
chiral sector of \none gauge theories, and specifically those with adjoint
matter.

For a theory with gauge group U($N$), and a superpotential 
tr\,${\cal W}(\Ph)$ for the adjoint chiral superfield $\Ph$,
the parameter space in question corresponds to the set of
dimensionless variables, modulo symmetries, that one 
can construct based on the parameters
in ${\cal W}(\Ph)$ and the dynamically generated SU($N$) scale $\La_{N=2}$. 
We will focus on the simplest nontrivial example, with 
a cubic superpotential,
\be
 {\cal W}_{\rm cl}(\Ph) =  \frac{1}{2} m\,\Ph^2 +
\frac{1}{3}g\,\Ph^3. \label{Wcl}
\ee
Classically this theory has 
two vacua within which the gauge group is unbroken, which is the 
phase on which we will focus. In the limit that $g=0$, the 
infrared theory is simply \none super Yang-Mills (SYM) and the dependence of ${\cal
W}|_v$ on $m/\La_{N=2}$ is fixed by a decoupling relation so that
only a single fractional instanton can contribute in the confining
vacua of the unbroken SU($N$). More generally, for finite $g$ one finds a 
nontrivial sum of fractional instanton contributions, leading to 
an effective superpotential of the form \cite{f2},
\be
 {\cal W}_{k} = \frac{2N}{3\la} \La_{N=1}^3\left( 1 \pm \left[1-\la
e^{2\pi i k/N}\right]^{3/2} \right), \;\;\;\;\;\;\; k=0,\ldots, N-1, \label{Weff}
\ee
in terms of the dimensionless parameter,
\be
 \la = \frac{8g^2\La_{N=2}^2}{m^2},
\ee
which is a convenient coordinate \cite{f2} for the parameter
space associated with the confining vacua (\ref{Weff}). 

These vacua lie on the maximally confining branch where the only
remaining massless field is a decoupled U(1) multiplet. However, this branch
connects to others associated with classical vacua for which
the gauge group is partially broken. It has
been conjectured that all these transitions are of two basic 
types \cite{f2}:
(1) where there are additional massless monopoles, such as the 
Seiberg-Witten singularities connecting Coulomb and Higgs branches
in \ntwo SYM; and (2) where the gluino condensate vanishes
corresponding to branch points in (\ref{Weff}). These two cases may alternatively
be characterised as the result of summation over instanton and
respectively fractional-instanton contributions. In the case 
of (\ref{Weff}), the branch points at \cite{ad,f2}
\be
 \la_k = e^{-2\pi i k/N}
\ee
also imply, for $N$ even, the presence of massless monopoles, and thus 
play a dual role. These points were first noted in the latter context,
namely as singular points connecting Coulomb and confining branches,
by Argyres and Douglas for SU(3) \cite{ad}. Furthermore, in
addition to the presence of these singularities connecting branches 
in different phases, it was shown by Cachazo, Seiberg and Witten
\cite{csw1} that particular phases may also exhibit multiple classical
limits, connected by smooth transitions through strong coupling regions
in parameter space. For the cubic model (\ref{Wcl}), these smooth
transitions always correspond to massless branches with $N\geq 4$.

The presence of a connected set of vacuum branches, in
general describing a multi-sheeted covering of the quantum parameter space, 
suggests that it may be profitable to study the action of symmetries
on this space. To this end, the aim of this paper is to
go beyond the vacuum structure and explore the spectrum of 1/2-BPS
states within the parameter space. For the \none theories studied
here, these BPS states are domain walls connecting discrete massive 
vacua, for which the data which can be extracted from the chiral
sector of the theory is limited to the tension, determined immediately
via the vacuum condensates, and also the multiplicity. The latter
degeneracy is formally determined by the CFIV index \cite{cfiv}
in the dimensionally reduced 1+1D theory obtained by compactifying
the worldvolume dimensions of the wall on a 2-torus. For our purposes, the CFIV
index will be understood to provide a definition of the 
wall multiplicity.

To determine the multiplicity of BPS states we require a Wilsonian
low energy description, and since the vacua of interest are confining
this requires some deformation of the original theory. We find that
compactifying the theory on a circle of radius $R$, and using the
known relation to integrable systems \cite{gorsky,dw}, 
is useful for this purpose. This approach also
provides a straightforward means of reproducing the vacuum 
structure, as first utilised by Dorey for the \nones theory
\cite{dorey}, and used recently by Boels et al. \cite{deB}
to determine the vacuum condensates for \none models with adjoint
fields, such as the example studied here.

We explore the multiplicity of certain BPS walls connecting confining
vacua as a function of the relevant parameter $\la$, and find that
the spectrum exhibits discontinuities on curves of marginal stability
(CMS), with certain bound states excluded from compact domains in
the $\la$-plane. The data describing the multiplicities of BPS states
thus form nontrivial sections over the parameter space, in a rather
close analogy to the way BPS particle multiplets form sections, 
transforming under subgroups of SL(2,$\Z$), over the moduli space of 
\ntwo gauge theories. While we do not yet see evidence for nontrivial
quantum symmetries of the latter form, we hope that further analysis
in this direction will lead to novel constraints.

The plan of the paper is as follows. In the next section, we describe
the compactification of the theory on $\R^3 \times S^1$, which 
allows a low energy effective superpotential to be obtained. This 
arises from instantons and thus has a natural semiclassical
interpretation. Using this effective theory, we reproduce the
condensates (\ref{Weff}) in the maximally confining vacua. 
In sections~3, we turn to BPS walls and study the structure of the
central charges as a function of $\la$, and exhibit a class of
marginal stability curves for composite states. The structure of these
curve simplifies for large $N$ and has an interesting connection with certain
singularities within the relevant Dijkgraaf-Vafa matrix model, which
we also describe. In Section~4, we turn to the dynamical question of
whether these marginal stability curves do indeed signify
discontinuities in the spectrum, and verify that this is the
case by first deducing the spectrum at a special point -- $\la=0$ --
and then constructing the leading order inter-wall potential as a
function of $\la$. This leads to the conclusion that specific composite
states are removed from the spectrum within a compact domain of the
$\la$-plane. Section~6 contains some concluding remarks, along with
further comments on the embedding of this theory within \nones SYM.

\section{Compactification and the Vacuum Structure}

The class of theories we will focus on here contains an \none
vector multiplet with gauge group U($N$) and an adjoint chiral 
multiplet $\Ph$, with scalar component $\ph$. The cubic superpotential
\be
 {\rm tr}\,{\cal W}_{\rm cl}(\Ph) = 
 \frac{1}{2} m\,{\rm tr}\,\Ph^2 + \frac{1}{3} g\,{\rm tr}\,\Ph^3
  \label{Wcl2}
\ee
leads to a set of classical vacua, distinguished by the distribution
of the $N$ eigenvalues $\varphi^a$ of $\ph$ between the two minima,
\be
 \varphi_{(1)} = 0, \;\;\;\;\;\; {\rm and} \;\;\;\;\;\; 
 \varphi_{(2)}=-\frac{m}{g}.
\ee
With $N_1$ eigenvalues equal to $\varphi_{(1)}$ and $N_2$ equal to
$\varphi_{(2)}$, satisfying $N_1+N_2=N$, the gauge group is broken
to U($N_1$)$\times$U($N_2$). We will be concerned primarily with 
those vacua where the gauge group is classically unbroken, i.e. with 
$(N_1,N_2)$ equal to $(N,0)$ and $(0,N)$. 
Note that the overall U(1) factor of the gauge group is central
and thus decouples as all fields are correspondingly uncharged.

Our aim in this section will be to construct a low energy effective
superpotential suitable for use in extracting the BPS
spectrum. This will necessarily involve a deformation of the theory. 
However, to first deduce the quantum vacuum structure on the
confining branch, we can proceed more directly. Treating
this system as a perturbation of \ntwo SYM, the vacuum structure
follows from the Seiberg-Witten solution. i.e.
given ${\cal W}_{\rm cl}$, we can write
\be
 {\cal W}_{\rm eff} = 
   \langle {\rm tr} {\cal W}_{\rm cl}(p\,\Id + \hat{\Phi}) \rangle
  = N {\cal W}_{\rm cl}(p) + \frac{1}{2}{\cal W}^{(2)}_{\rm cl}(p) 
    \langle {\rm tr} \hat{\Phi}^2 \rangle, \label{weff1}
\ee 
where $\hat{\Ph}$ transforms in the adjoint of SU($N$), and so we have
made use of the relations $\langle {\rm tr} \hat{\Phi}^{2n+1}
\rangle=0$, while
$p=\langle {\rm tr}(\Ph)\rangle/N$. For confining vacua, the maximal degeneration 
of the SU($N$) Seiberg-Witten curve \cite{af,klyt} leads to the non-vanishing 
condensate \cite{ds} 
$\langle {\rm tr} \hat{\Phi}^{2} \rangle = 2N \om_k\La_{N=2}^2$ 
with  $\om_k=\exp(2\pi i k/N)$ an $N^{th}$-root of unity. On
substitution into (\ref{weff1}), one finds \cite{f2,cdsw}
\be
 {\cal W}_k(p) =N \left[{\cal W}_{\rm cl}(p) + m(p)\om_k\La_{N=2}^2 \right],
  \;\;\;\;\;\;\; k=0,\ldots,N-1, \label{Wp}
\ee
where $m(p)={\cal W}_{\rm cl}''(p)$ is the effective mass term for all
the scalar modes, including the trace component $p$, at weak gauge coupling.
For the cubic case this result, first obtained in \cite{f2,cdsw}, has 
the simple interpretation of contributing the gluino condensate
from the confining SU($N$) factor. More generally, one expects a 
nontrivial (but finite) sum of such fractional 
instanton contributions\footnote{In certain cases, there may also be
  additional instanton contributions, whose form
  determines the ultraviolet completion of the theory. See
  \cite{aivw} for a nice discussion of this issue.}. Integrating
out the trace component $p$, we recover the confining vacua
(\ref{Weff}) noted earlier.

When the trace component $p$ is light relative to the other modes --
and indeed it becomes massless when $\la=\la_k$ -- the effective
superpotential (\ref{Wp}) may provide a reliable low energy
description. However, since (\ref{Wp}) depends explicitly on the
vacuum, we cannot use it to study wall configurations which
interpolate between different vacua. These configurations must
necessarily excite the gauge modes responsible for gluino
condensation, and we will need to include them in any tractable low
energy description. Moreover, since the vacua of interest are confining, some
deformation of the theory is required. To this end, if we wish to avoid adding
additional fields, a natural approach is to compactify the theory
on a circle of radius $R$. A Wilson line for the gauge field then 
provides a massless scalar, which may be rotated to the Cartan
subalgebra $\rh_a = \int_{S^1} A_a$,
$a=1\ldots N$, and this will generically Higgs the gauge group to its 
maximal torus. The resulting
photons can be dualised to periodic scalars $\si^a$ \cite{polyakov},
and supersymmetry then ensures that the complex combination,
\be
 \tilde{q}_a = \rh_a + \ta \si_a,
  \;\;\;\;\;\;\;\; a=1,\ldots,N, \label{qdef}
\ee
where $\ta$ is the complex gauge coupling, is the lowest component
of a (classically massless) chiral superfield which characterises the gauge sector of the
dimensionally reduced low energy theory. The additional adjoint scalars 
reduce straightforwardly, and the effective 3D theory is a 
Wess-Zumino model with two adjoint chiral fields and a classical
superpotential given by (\ref{Wcl2}). We now turn to possible
nonperturbative corrections.

\subsection{Semiclassical interpretation of ${\cal W}_{\rm eff}$}

When the compactification radius $R$ is sufficiently small, 
$\La R \ll 1$, the system is weakly coupled and thus any
nonperturbative corrections to ${\cal W}$ should have a
semiclassical interpretation in terms of instantons. Therefore, we can 
parametrise the superpotential in the form
\be
 {\cal W}_{\rm eff} = {\cal W}_{\rm cl}(p_a) + {\cal W}_{\rm
inst}(q_a,p_a), \label{Wsemi}
\ee
where ${\cal W}_{\rm class}(p_a)$ is the classical superpotential 
(\ref{Wcl2}) and we have now chosen to denote the eigenvalues of the 
adjoint field $\Ph$ as $\{p_a\}$, $a=1,\ldots,N$.

The form of the instanton generated superpotential in these theories
was first explored by Polyakov \cite{polyakov}, and for theories
with \ntwo SUSY by Affleck, Harvey and Witten \cite{ahw}. For
pure \none SYM, there are $N$ nontrivial 1-instanton configurations,
corresponding to $N-1$ BPS monopoles aligned along simple roots and,
for finite $R$, an additional `Kaluza-Klein'-monopole wrapped
around the compact direction \cite{ly}, which 
contribute to the 
superpotential \cite{sw96,kv,hg,ahiss,bho,dorey,dkm,dhk,ritz}. 
In the presence of additional 
adjoint chiral fields, these configurations would normally have too
many fermionic zero modes to contribute to ${\cal W}$, but if these fields are
classically massive, the additional zero modes can be absorbed by
the mass terms. Accounting for the monopole-antimonopole background
\cite{polyakov} which resums the instanton vertices to a
field-dependent superpotential, we can write down its generic form
as follows \cite{dkm,dhk}\footnote{For simplicity, we
will write the superpotential using a 3+1D normalisation in what
follows, so that ${\cal W}_{\rm 3D} = 2\pi R\, {\cal W}_{\rm eff}$.},
\be
 {\cal W}_{\rm inst} = \frac{1}{2}\La_{N=2}^2
   \sum_{a=1}^N [m( p_a )+m( p_{a+1})] 
   e^{q_a - q_{a+1}}, 
  \;\;\;\;\;\; m(p_a) = m + 2 g p_a, \label{Wsemi2}
\ee
in which $m(p_a)$ is the classical mass term for $\varphi_a$ given
by expanding ${\cal W}_{\rm cl}$ about a given vacuum $\langle p_a
\rangle$. Note that 
$q_a$ and $\tilde{q}_a$ in (\ref{qdef}) are related by a constant
shift depending on the coupling $\ta$. The structure of the mass
insertions follows from the choice of basis for the scalars which
corresponds to taking projections along the fundamental weights. 
The monopoles in contrast are aligned along the simple roots, and 
thus are embedded along two fundamental weight directions.

In the expression (\ref{Wsemi2}), we have slightly extended
the result of Davies et al. \cite{dkm,dhk}, having 
made the replacement 
$\langle p_a \rangle \rightarrow p_a = \langle p_a \rangle + \de p_a$,
so that a genuine interaction term is introduced. We have not derived
the appropriate instanton vertex explicitly, as it can be justified 
in a different manner to be discussed shortly. In particular,
for later use, it is convenient to recall another approach to this 
problem which starts from the \ntwo limit and uses the relation to integrable
systems \cite{dorey}.

\subsection{The effective Toda superpotential}

At a formal level, the calculation of ${\cal W}_{\rm eff}$ is
dramatically simplified by recalling the relation between
the Seiberg-Witten solution for \ntwo SYM and certain
(complexified) integrable systems \cite{gorsky,dw}. More precisely, written 
in terms of the natural gauge invariant coordinates 
on the \ntwo moduli space,
\be
 u_n \equiv \frac{1}{n}{\rm tr}\, \ph^n,
\ee
the superpotential takes the form
\be
 {\rm tr}\,{\cal W} = m u_2 + g u_3, \label{sp}
\ee
where we have now dropped the subscript as one can argue that this
form is exact within the compactified theory on $\R^3 \times
S^1$. To do this, one treats this superpotential
as a perturbation of the (reduced) \nfour theory, and interprets the coordinates
$u_n=u_n(x)$ as constrained variables depending on the underlying
Seiberg-Witten curve. To make use of this expression one must
find the corresponding unconstrained variables parametrising the
Jacobian of the curve.

This problem is elegantly solved via the correspondence with
integrable systems, and in this case the affine Toda lattice
\cite{gorsky}. This
relation is perhaps most transparent on noting that the Seiberg-Witten
curve for U($N$) can be written in the form,
\be
 P_N(x,u_n) = z +\frac{\La^{2N}_{N=2}}{z}, 
 \;\;\;\;\;\;\;\;\;\; P_N(x,u_n)= x^N + \sum_{n=1}^N s_n x^{N-n},
\ee
where $s_n$ and $u_n$ are related by the recursion relation, $rs_r +
\sum_{i=0}^r s_{r-i}u_i =0$, with $u_0=0$ and $s_0=1$. This curve is 
identifiable as the spectral curve for the affine Toda lattice, namely
\be
 {\rm det}\, (x\Id - L(z)) = 0,
\ee
where the Lax matrix is given by
\be
 L(z) = \left( \begin{array}{ccccc} p_1 & \La_{N=2}\,e^{q_1-q_2} & 0 & \cdots & z\\
                          1 & p_2 & \La_{N=2}\,e^{q_2-q_3} & \cdots & 0\\
                          0 & 1 & p_3 & \cdots & 0 \\
                          \vdots & \vdots & \ddots & \ddots & \vdots
                          \\
                      z^{-1}\La_{N=2}\,e^{q_N-q_1} & 0 & \cdots & 1 &
                      p_N\end{array}\right)
\ee
in terms of the canonical variables $\{p_a,q_a\}$. 
Thus the unconstrained variables on the Jacobian of the curve are precisely 
the (complex) coordinates and momenta $\{q_a,p_a\}$ for the affine Toda lattice.
The conserved quantities, to be identified with $u_n$, are given
by
\be
 u_n = \frac{1}{n} {\rm tr}\, L^n,
\ee
which allows us to identify the momenta $p_a$ with the
eigenvalues of the adjoint scalar $p_a=\langle \varphi_a \rangle$,
explaining the notation introduced above.

With this parametrisation, the superpotential (\ref{sp}) takes the form
\be
 {\cal W}_{\rm eff} = \frac{1}{2}m\left[ \sum_{a=1}^N p_a^2 +
 2\La_{N=2}^2 e^{q_a-q_{a+1}}\right]
 + \frac{1}{3} g \left[ \sum_{a=1}^N p_a^3 +
3\La_{N=2}^2 (p_a + p_{a+1}) e^{q_a-q_{a+1}}\right]  \label{spf}
\ee
which, on rewriting it in the form (\ref{Wsemi}), agrees with
the expected 1-instanton corrected potential\footnote{Note that if we were to
consider a classical superpotential of degree four or higher, 
e.g. ${\rm tr}\,\Ph^4$, then
the semiclassical interpretation would require the contribution 
of certain multi-instanton corrections. It would be interesting to see
precisely how the additional zero modes are lifted in such cases.}.

This latter approach for calculating ${\cal W}_{\rm eff}$, making use
of the connection to integrable systems, was first used by Dorey
for the \nones theory, and has since been applied to other
theories including pure \none SYM \cite{ritz,dhk} and more
general deformations ${\cal W}_{\rm cl}$ \cite{ds2,dhks}. A wider class
of U($N$) examples involving polynomial superpotentials was also 
recently studied in this way by Boels et al. \cite{deB}, and
subsequently extended to other gauge groups \cite{ogg} .

\subsection{Quantum vacuum structure}

Extremising the superpotential (\ref{spf}), we find the following
equations,
\bea
 p_a(\et p_a+1) &=& = -\et (x_a + x_{a+1})  \label{eom1}\\
 \left(1+\et (p_a+p_{a+1})\right) x_a &=& 
 \left(1+\et (p_{a-1}+p_{a})\right) x_{a-1} \label{eom2},  
\eea
where $\et=g/m$ and $x_a=\exp(q_a-q_{a+1})$, and we have
imposed the decoupling condition $q_{N+1}=q_1$ for the overall
U(1) factor. Since we are interested in vacua with a classically
unbroken gauge group, we solve (\ref{eom2}) by setting
\be
 q_a = q, \;\;\;\;\; p_a=p, \;\;\; \forall a=1,\ldots,N.
\ee
The resulting $2N$ vacua are given by 
\be
 u_1 = \sum_a p_a = N p^{\pm}_k 
  = \frac{N}{2\et} \left( 1 \mp \sqrt{1-\la x_k}\right), 
\ee
with
\be
 x_k = e^{2\pi i k/N}, \;\;\;\;\;\; k=0,\ldots,N-1,
\ee
where $u_1$, and thus $p$, is single valued on a
double-sheeted cover of the $\la$-plane, and the two sheets are 
distinguished by the corresponding classical vacuum expectation value 
for $\ph$. One may note here that the intuitive correspondence
between massive vacua and (complex) mechanical equilibria of the integrable
system, that is so apparent for SU($N$) \cite{dorey}, appears to
have been lost here due to the nontrivial vacuum expectation value for $p$,
so that the `equilibria' now have nonzero coordinate
momentum. However, one finds that these vacua still correspond to
stationary points where the angular momentum
within the Jacobian vanishes \cite{holl2,dB2}. In other words, there
is a canonical transformation within the integrable system to an
action-angle basis for which the massive vacua still 
map to mechanical equilibria. 

The $2N$ extrema of the superpotential are then given by,
\be
 {\cal W}^{\pm}_k = \frac{2N}{3\la}\La_{N=1}^3 
 \left( 1 \mp [1-\la x_k]^{3/2}\right),
 \;\;\;\;\;\;\;\; k=0,\ldots,N-1, \label{W0}
\ee
and these results are fully consistent with those recently 
obtained using Seiberg-Witten and matrix model 
techniques \cite{f2,cdsw} as discussed above.

\EPSFIGURE[t]{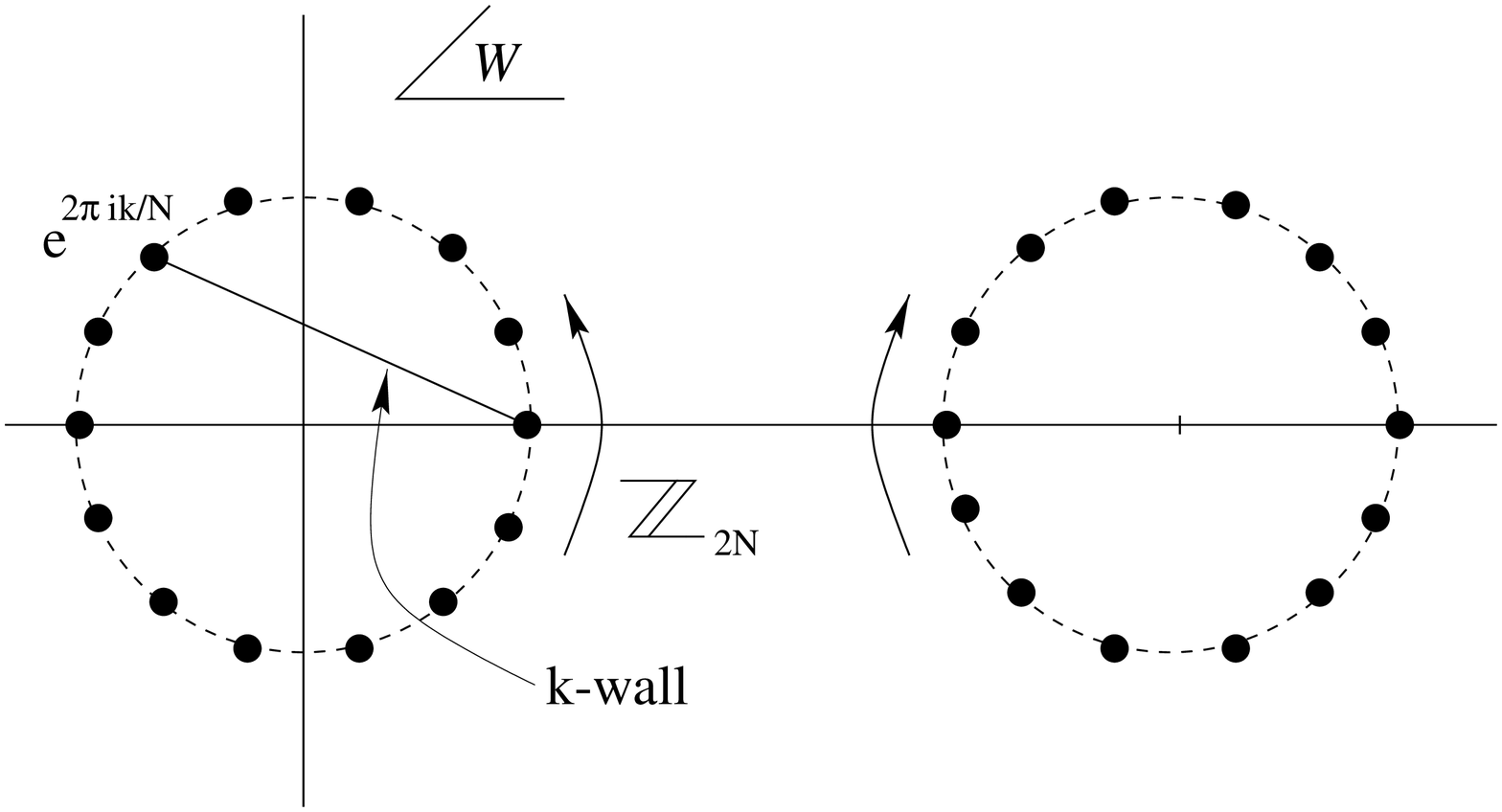,height=5cm}{\footnotesize Confining 
   vacuum structure 
   in the ${\cal W}$-plane for $N=14$ and $\la \ll 1$, exhibiting the
   action of the cyclic `$\Z_{2N}$ symmetry', and illustrating the 
   interpolating profile of a $k$-wall.} 

For $\la \ll 1$ the vacua fall on two approximately circular curves
in the ${\cal W}$-plane, as exhibited in Fig.~1. For the vacua 
corresponding to the classical vacuum at $\phi=0$, the asymptotics 
near $\la=0$ have the form
\be
 {\cal W}^+_k(\la\rightarrow 0) \sim N \La_{N=1}^3e^{2\pi i
k/N}\left( 1 + {\cal O}\left(g^2
\left(\frac{\La_{N=1}}{m}\right)^3\right)\right),
\ee
where the leading term corresponds to the pure \none SYM limit, and 
the subleading terms, proportional to powers of $e^{2\pi i \ta(m)/N}$, 
correspond to higher order fractional instantons.

The plot also exhibits a discrete cyclic symmetry which rotates
the vacua via its action on $\la$,
\be
 \la \rightarrow e^{2\pi i/N} \la, \label{rotate}
\ee
which we interpret as induced by SL(2,$\Z$) translations on the bare gauge
coupling $\ta \rightarrow \ta + 1$. The corresponding action on 
the vacua within each branch is then
\be
 {\cal W}^{\pm}_k \rightarrow {\cal W}^{\pm}_{k+1},
\ee
while the rotation of $\la$ around one of the branch points
induces the action \cite{f2}
\be
 {\cal W}^{\pm}_k \rightarrow {\cal W}^{\mp}_{k}.
\ee
With a suitable choice of branch, the action (\ref{rotate}) rotates
${\cal W}_k$ through all $2N$ vacua on both branches, and we can interpret this
`$\Z_{2N}$ cyclic symmetry' as a nontrivial extension of the $\Z_N$
nonanomalous discrete subgroup of the classical U(1)$_R$ symmetry,
which rotates the vacua of pure SYM. The extension arises directly
from the double-sheeted structure of the confining branch over the
quantum parameter space, coordinatised by $\la$. 

Note that this structure does not, however, extend globally over the parameter
space. In particular, the asymptotics near $\la=\infty$ are
\be
 {\cal W}^{\pm}_k(\la\rightarrow \infty) \sim \pm
\frac{4\sqrt{2} i}{3} N g \La_{N=2}^3 e^{3\pi i k/N} \left( 1 +  {\cal
O}\left(\frac{1}{g^2}\left(\frac{m}{\La_{N=2}}\right)^2\right)\right).
\ee
and $\la$ is no longer a good coordinate on the parameter space --
the leading dependence is instead on $\sqrt{\la}$.

We have focused on the maximally confining vacua, but for
completeness we note that for generic $N$ there are in addition a
large number of branches with
massless vacua which map to classical vacuum solutions with a broken gauge
group in the limit that $\La_{N=2}\rightarrow 0$. The cubic 
superpotential only exhibits classical vacua where the
the eigenvalues $\{p_a\}$ are split into two groups, and thus 
there is a single (nontrivial) massless U(1) factor.
The structure of these vacua was elaborated for $N\leq 6$
in \cite{csw1}, and shown to include branches where smooth transitions
in parameter space between distinct classical limits were possible.
These vacua decouple in the limit that $\la\rightarrow 0$, where
the theory reduces in the infrared to pure SYM. However, in the
opposite limit, $\la \rightarrow \infty$, certain massless vacua survive
and in the U(3) case, for example, apparently asymptote to the Argyres-Douglas
conformal points \cite{ad}.

\section{BPS Wall Kinematics}

With a clear picture of the confining vacuum structure in the ${\cal
W}$-plane, we can now turn to the question of the spectrum of
BPS states, namely domain walls interpolating between distinct vacua.
Given that the vacua described above are distinct, such
configurations are necessarily present on topological
grounds. However, the number of corresponding BPS states is a 
dynamical question that we will come to shortly. We first outline the 
kinematic structure imposed by the \none superalgebra.

\subsection{Central Charges}

The vacua labelled by ${\cal W}_k$ allow us to isolate
two sets of walls, characterised by whether or not both vacua at
spatial plus and minus infinity are on the same branch. If they are,
then such walls are present only at the quantum level. However, if the
asymptotic vacua are on different branches then the corresponding 
walls are visible classically. More precisely, as noted in \cite{f2},
this characterisation is useful for $\la \ll 1$ while, as we will see, 
by varying $\la$ one observes various discontinuities in the spectrum
of BPS walls, with the underlying reason being the presence of curves
of marginal stability in the parameter space coordinatised by $\la$. 

We begin by discussing the central charge structure, and to this end
it is useful to introduce the following (condensed) notation for the 
vacua,
\be
 {\cal W}_k = \frac{2N}{3\la}\La_{N=1}^3 
 \left( 1 - [1-\la x_k]^{3/2}\right), \;\;\;\;\; k=0,\ldots,2N-1,
\ee
which incorporates both branches, with the implicit understanding that
the vacua for $k=N,\ldots,2N-1$ lie on the second branch. This
notation is consistent with the discrete $\Z_{2N}$ symmetry noted above.
It is then convenient to introduce the (SU($N$) root-valued) 
topological charges,
\be
 n^k_{(ij)}= \de^k_j - \de^k_i,
\ee
in terms of which the central charges for walls interpolating 
between vacua labelled by $i$ and $j$ phase-units respectively 
are given by \cite{ds3}
\be
  {\cal Z}_{ij} = 2 n^k_{(ij)}\left[ {\cal W}_k -\frac{1}{16\pi^2}\left( T_G - \sum_f
      T(R_f)\right) \langle {\rm tr} W^{\al}W_{\al}\rangle_k \right].
\ee
The second term here is an anomaly, which vanishes in the present case
due to the \ntwo matter content. Consequently,
we have 
\be
 {\cal Z}_{ij} = 2n^k_{(ij)}{\cal W}_k = 2\left({\cal W}_j - {\cal W}_i\right),
\ee
or more explicitly, 
\be
 {\cal Z}_{jk}
    = \frac{4N}{3\la} \La_{N=1}^3\left[ 
 \left(1-\la e^{-2\pi i k/N}\right)^{3/2} 
  - \left(1-\la e^{-2\pi i j/N}\right)^{3/2} \right],
   \label{Zjk}
\ee
while the corresponding wall tensions are 
\be
 T_{jk} = |{\cal Z}_{jk}|.
\ee
Note that $T_{kk+N}$ goes to zero at the branch points in 
the $\la$-plane \cite{f2}, corresponding to the fact that the vacua
labelled by ${\cal W}_k$ and ${\cal W}_{k+N}$ collide at these points.

\subsection{Marginal Stability}

We now specialise to walls interpolating between two vacua on the same
branch, and for definiteness restrict to walls associated with the
central charges ${\cal Z}_{-kk}$. For sufficiently small $\la$, such 
configurations are naturally interpreted as bound states of $2k$
minimal 1-walls with charges 
${\cal Z}_{pp+1}$ which interpolate between adjacent vacua.
For this reason, we will generally focus on the simplest bound state
${\cal Z}_{-11}$ composed of two 1-walls. Supersymmetry
demands that such putative BPS bound states, when present, are at
least marginally bound. The corresponding submanifolds on which
\be
 T_{-11} = T_{-10} + T_{01} \label{Tcms}
\ee
are of co-dimension one in the parameter space, and allow for possible
discontinuities in the spectrum, where BPS bound states may delocalise
and leave the spectrum. The position of these curves of marginal
stability (CMS) in parameter space is fixed by supersymmetry and
therefore is purely kinematic. It is this question that we will
address in the remainder of this section, while the subsequent
dynamical issue of whether discontinuities in the spectrum do actually
occur on these submanifolds will be addressed subsequently.

A convenient characterisation of the submanifolds on which
(\ref{Tcms}) holds is that the relative phase of the two constituent
central charges vanishes,
\be
 \om(\la)|_{\rm CMS} = 0, \label{cmsdef}
\ee
where 
\be
 e^{i\om} \equiv \frac{{\cal Z}_{-10} 
    \bar{{\cal Z}}_{01}}{|{\cal Z}_{-10}{\cal Z}_{01}|}. \label{omdef}
\ee
Using the explicit form for the central charges (\ref{Zjk}), the 
condition (\ref{cmsdef}) implies the following two (real) constraints:
\be
 {\rm Im}(f_1(\la)f_1(\bar\la)) = 0, \;\;\;\;\;\;\;\; 
 {\rm sgn}(f_1(\la)f_1(\bar\la)) < 0, \label{cmsexp}
\ee
where 
\be
 f_k(\la) \equiv 1 - \left(1-2i\frac{\la}{1-\la}e^{i\pi k/N} 
 \sin \frac{\pi k}{N}\right)^{3/2}.
\ee
The CMS defined by the constraints in (\ref{cmsexp}) is exhibited 
in Fig.~2 for several values of $N$; it defines a 
closed curve for $N>6$, and has a simple limiting form for large $N$
that we will now describe in more detail.

\FIGURE[t]
  {\parbox{12cm}{
    \centerline{%
   \psfig{file=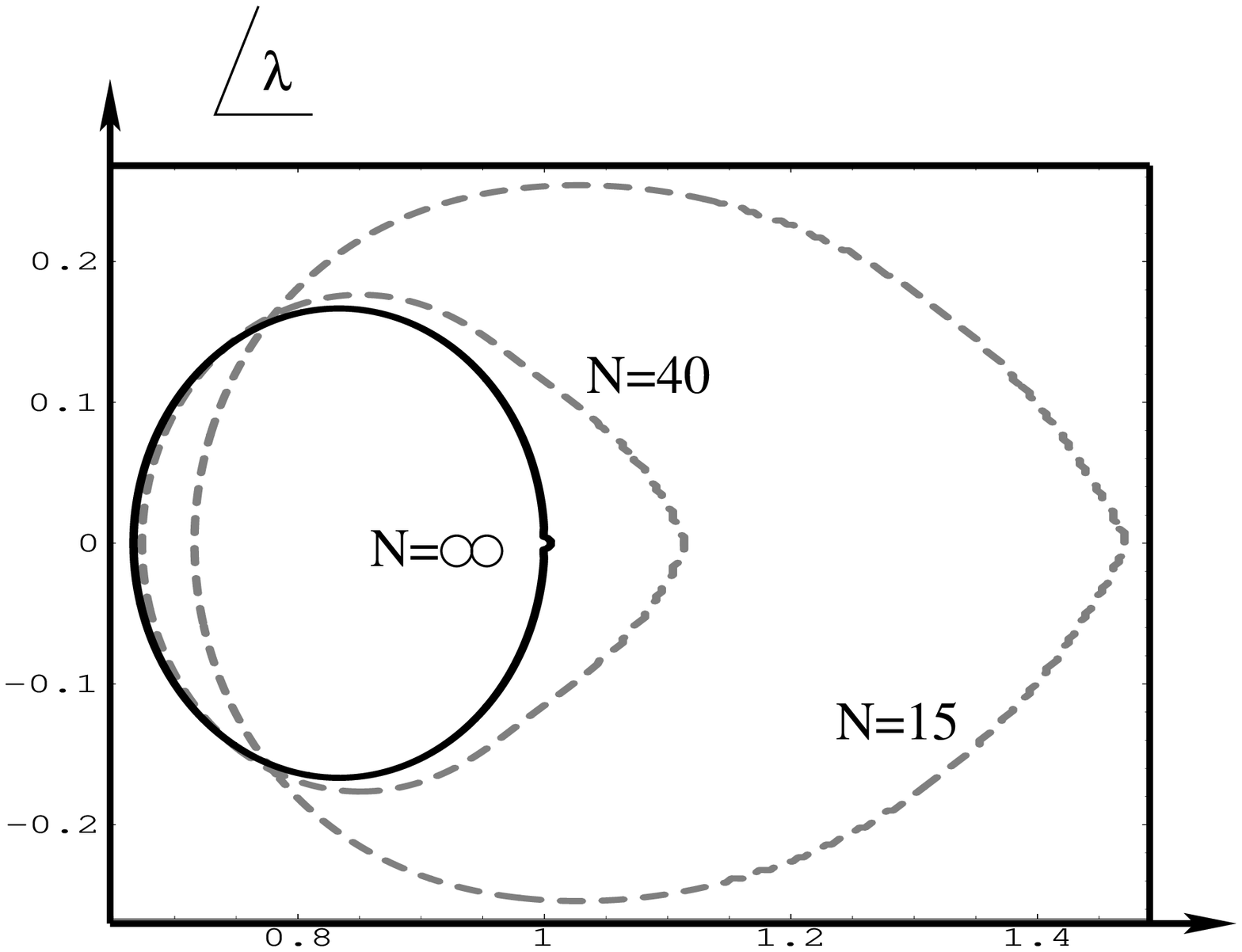,width=9cm,angle=0}%
         }
\newcaption{\footnotesize A plot of the CMS curves in the $\la$-plane 
  for the 2-wall composite, and several values of $N$. For
  $N\rightarrow \infty$ the limiting form of the curve is a circle
  described by (\ref{laph})}}}

As follows from (\ref{Zjk}), the expression (\ref{omdef}) for the
relative phase contains branch points at $\la = \{e^{-2\pi i/N}, 1,
e^{2\pi i/N}\}$ internal to the $\la$-plane, and this branch
structure is inherited by the equation (\ref{cmsexp}) defining the
CMS. Although having physical significance in signalling the presence
of singular points connecting this branch with others associated with
massless  vacua, these points are not crucial here in that the spectrum of finite tension
walls is at most discontinuous on curves of co-dimension one, and so
the co-dimension two branch points can always be avoided.

To make use of this fact, we can simplify the structure of the CMS
by taking $N$ large, in a scaling limit such that
\be
 N\om(\la)\, = \,{\rm constant},
\ee
whereupon at leading nontrivial order the
branch points lie along a line, $\la_k \sim 1 \pm 2\pi i k/N$.
In this regime it is straightforward to evaluate the 
relative phase $\om(\la)$ for the ${\cal Z}_{-11}$ wall bound state. 
We obtain the following simple expression,
\be 
 \om(\la) = \frac{2\pi}{N} \left[ 1- \frac{1}{2} {\rm Re}
  \left( \frac{\la}{1-\la}\right)\right] 
   +{\cal O}\left(\frac{1}{N^{2}}\right),
\ee
which reduces to $\om=2\pi/N$ in the limit $\la \rightarrow 0$ of
pure SYM, and more generally is a harmonic function of $\la$,
with a corresponding pole at the branch point $\la_0 = 1$ associated
with the intermediate vacuum.

From this expression, we find that at large $N$ the phase $\om$ vanishes on a 
circle in the complex $\la$-plane described by 
\be
 {\rm CMS:} \;\;\;\; \left| 6\la-5 \right|^2 = 1. \label{laph}
\ee
Note that since the curve only touches the line $\la=1+i\al$ at the
point $\al=0$, it is consistent to track the curve all
the way to $\la=1$ since it avoids the additional branch points at
$\la_{-1}$ and $\la_{1}$. There are apparently two distinguished
points on the CMS, at $\la=1$ and $\la=2/3$, where the curve
intersects the real axis. The former corresponds to the branch
point for the intermediate vacuum, while the latter has an interesting
interpretation within the Dijkgraaf-Vafa matrix model \cite{dv}, that we will
explore in more detail below.

\subsection{The matrix model}

The particularly simple large-$N$ structure for the CMS determined above
hints at a more transparent interpretation. One point of view that we
will provide some evidence for below is that the onset of marginal stability
in this regime is characterised by a qualitative change of the
intermediate vacuum. For the composite state considered above, the 
relevant vacuum has a critical point at $\la=1$,
and presumably also intersects with massless vacuum branches at other
points whose precise locations depend on $N$ \cite{f2,cdsw}.

The fact that the CMS passes through the critical point at $\la=1$ is
then not too surprising as the intermediate vacuum at this point will
have additional massless excitations. The second intersection of the CMS with
the real axis at $\la=2/3$ is less immediately attributable to any
pathology of the vacuum. However, it is a distinguished point within
the geometric description of the vacua in the \ntwo limit, and thus
also within the matrix model picture.

To illustrate this point, we first recall some of the relevant aspects
of the Dijkgraaf-Vafa matrix model. In particular, for the theory at hand,
there now exist some rather general arguments \cite{dv,dglvz,cdsw}
implying that specific 
aspects of the chiral sector of the theory 
are captured by a holomorphic matrix model of the form
\be
 e^{-M^2 {\cal F}(S)/S^2} 
  = \int [d X] e^{-\frac{M}{S} {\rm tr}\,{\cal W}_{\rm cl}(X)},
\ee
where $X$ is an $M\times M$ matrix, and we take $M\rightarrow \infty$
to isolate the planar sector. $S$ serves as a loop counting parameter in the planar
limit of the matrix model, but ultimately, one makes the identification
$S = - \frac{1}{32\pi^2}{\rm tr} \, W^{\al}W_{\al}$ within the gauge
theory. The large-$M$ saddle point is 
characterised by the condensation of the eigenvalues of $X$ into a set of
cuts $[a_k,b_k]$, each corresponding classically to one of the
roots of ${\cal W}_{\rm cl}'(x)=0$. The saddle point condition is
conveniently expressed in terms of the force on a test eigenvalue,
\be
 y_{\rm m}(x) = {\cal W}'_{\rm cl} - 2 S {\cal R}_m \;\;\;\;\; {\rm
where} \;\;\;\;\;
 {\cal R}_{\rm m} = \frac{1}{M} \left<{\rm Tr}\,\frac{1}{x-X}\right>. \label{resolv}
\ee
${\cal R}_{\rm m}$ is the trace of the resolvent whose discontinuity across
the cuts determines the eigenvalue distribution (see e.g. \cite{gzj}). 
We are concerned here with 1-cut solutions, corresponding to
classically unbroken U($N$) vacua, and in 
terms of $y_m$ the (large $M$) saddle point requires \cite{dv,f2},
\be
 y_{\rm m}^2 = ({\cal W}'_{\rm cl}(x))^2 - f_{1}(x) = g^2(x-x_*)^2
(x-a)(x-b) \label{surf}
\ee
where $f_1$ is a polynomial of degree one, and $a$ and $b$ are the
endpoints of the cut, with $x_*$ a double point corresponding to
the degenerate second cut. This constraint describes a degenerate hyper-elliptic 
curve, and is equivalent to the appropriate degeneration of the
U($N$) Seiberg-Witten curve \cite{af,klyt}. 

For this saddle point configuration, the gauge theory
superpotential is determined by matrix model prepotential as \cite{dv}
\be
 {\cal W}_{\rm m} = N \Pi_B - (\ta+k) \Pi_A, \label{spmm}
\ee
in the 1-cut sector, where the periods $(\Pi_A,\Pi_B)$ are given 
in terms of the resolvent ${\cal R}_{\rm m}$ via
\be
 \Pi_A = 2\pi i S = \oint_A  y_{\rm m}, \;\;\;\;\;\;\;\;
 \Pi_B = 2\pi i \frac{\ptl {\cal F}}{\ptl S} = \oint_B  y_{\rm m}, \label{per}
\ee
using a symplectic basis of (compact) $A$ and (noncompact) 
$B$ cycles for the surface (\ref{surf}), as shown in Fig.~3.

\FIGURE[t]
  {\parbox{10cm}{
    \centerline{%
   \psfig{file=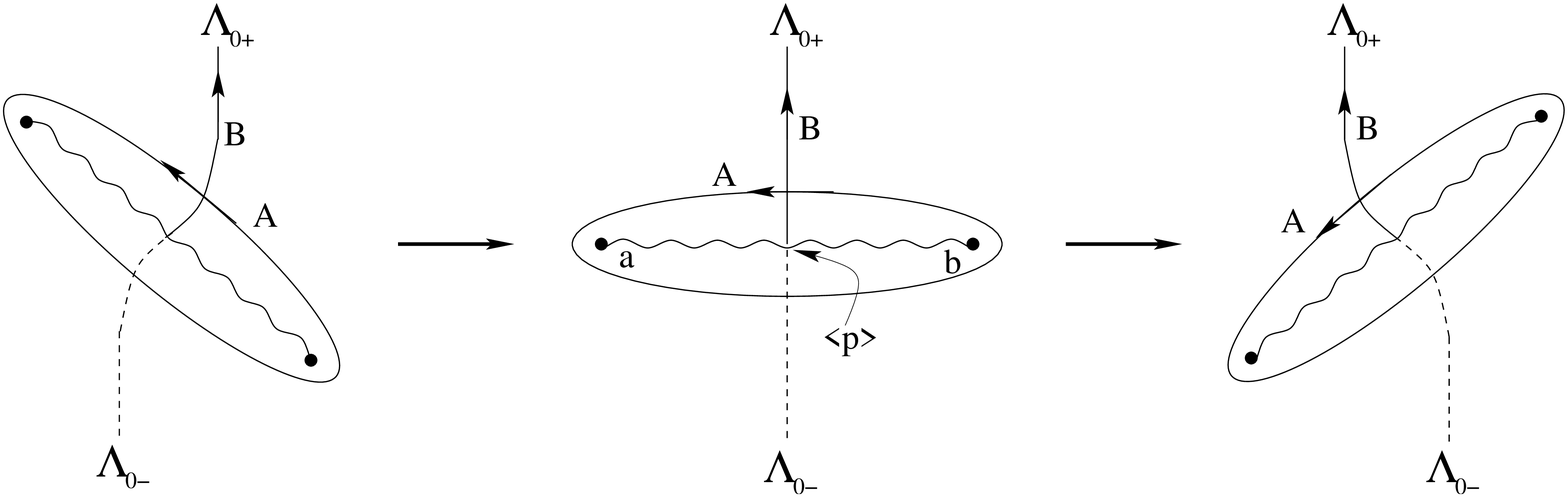,width=14cm,angle=0}%
         }
\newcaption{\footnotesize A conventional basis of $A$ and $B$ cycles
  for the 1-cut degeneration of the curve. The centre of the cut is
  identified with $p$, and the cut `rotates' as shown in
  passing from one vacuum to another via a domain wall.}}}

Making use of (\ref{surf}), the integrals in (\ref{per}) may be
expressed in terms of elementary functions. Evaluating $\Pi_B$ in the
limit $\La_0\rightarrow \infty$, and dropping an irrelevant constant, 
one finds that the contribution from $\Pi_A$ in (\ref{spmm}) serves to
reconstruct the dependence on the dynamical scale $\La_{N=2}$. With 
$p=p(S)$ denoting the mid-point of the cut, one finds that 
\cite{dv,f2,cdsw,civ},
\be
 {\cal W}_{\rm m}(S) = N{\cal W}_{\rm cl}(p(S)) 
 + S \ln \left[ \frac{e\,m(p(S))\La_{N=2}^2}{S}\right]^N. \label{WSmm}
\ee
This result --  a generalisation of the Veneziano-Yankielowicz
\cite{vy} superpotential -- is identical\footnote{The on-shell 
  equivalence between the `equilibria' of the integrable
  Toda system and the matrix model saddle points has recently been
  verified more generally \cite{holl2,dB2}.} to the answer one obtains from (\ref{Wp}) by
integrating in $S$ via a Legendre transform with respect to $\ln
\La^{2N}_{N=2}$ \cite{cdsw}, and then integrating out $p$ in favour of
$S$. On extremising ${\cal W}_{\rm m}(S)$, and with the implicit dependence on the 
scale $\La_{N=2}$, the on-shell superpotential takes the form,
\be
 {\cal W}_{\rm m}|_k = N \Pi_B(\ta_k),
\ee
where
\be
 \ta_k \equiv \frac{\ptl \Pi_B}{\ptl \Pi_A}  = \frac{\ta+k}{N} \label{tak}
\ee
is the modular parameter of the noncompact elliptic
curve, indicating that it is an $N$-fold cover of the bare curve. Of
course, a renormalisation group-invariant version of this statement is
that ${\cal W}_{\rm m}|_k$ depends on $N$ only via the 
(complexified) dynamical scale
\be 
 \La_{N=2}^2\exp\left(\frac{2\pi i k}{N}\right)
   = \La_0^2 \exp\left(\frac{2\pi i (\ta+k)}{N}\right).
\ee
This explanation of the remarkably simple $N$-dependence of the
vacuum values of the superpotential, and consequently of the central
charges is one of the virtues of the matrix model approach, since the 
matrix model itself is $N$-independent, and results from the 
planar saturation of the chiral sector \cite{dv}. In particular, using
the topological charges $n^k_{(ij)}= \de^k_j - \de^k_i$, the central
charges can be compactly written in terms of the $B$-period as follows,
\be
 {\cal Z}_{ij} = 2 n^k_{(ij)}{\cal W}_k = 
 2 N n^k_{(ij)} \oint_{B}y_m(\ta_k).
\ee
Geometrically, the 1-cut configuration is unique, with the nontrivial
vacuum structure arising via extremisation of (\ref{spmm}) for a given
value of $k$. However, for the purpose of interpreting domain walls in
this picture, it is useful to note that the rotation  of the cutoff
$\La_0 \rightarrow \La_0 e^{2\pi i}$ induces the monodromy
$\Pi_B \rightarrow \Pi_B - \Pi_A$ \cite{civ}, and thus if we choose to
fix the phase of the regulator $\La_0$, we can view the process of 
passing through a wall as corresponding to a rotation of the cut as
exhibited in Fig.~3, with the corresponding rearrangement of the 
eigenvalues\footnote{In the context of `classical walls', where
  the cuts corresponding to the two vacua have distinct classical
  limits, it was noted by Dijkgraaf and Vafa that walls are apparently 
  dual to eigenvalue tunnelling \cite{dv}. Such an `instanton-like'
  interpretation seems more problematic here as the relevant vacua are
  classically degenerate.}.

With this formalism in hand, we now reconsider the near-CMS regime.
and in particular the intercept points $\la=1$ and $\la=2/3$. As
follows from (\ref{surf}), both these points correspond to
degenerations of the 2-cut solution as shown in Fig.~4. In the
plot, for illustration, we show some of the 
additional cycles relevant within a 2-cut
degeneration of the Seiberg-Witten curve. However, in the 1-cut
solution we consider here, the only remnant of these degenerations
is that at $\la=1$, the double point at $x=x_*$ lies at the
mid-point of the cut, while at $\la=2/3$ the double point lies at
one end. The degenerating cycles do not play a direct role in the 
matrix model solution. 

\FIGURE[t]
  {\parbox{10cm}{
    \centerline{%
   \psfig{file=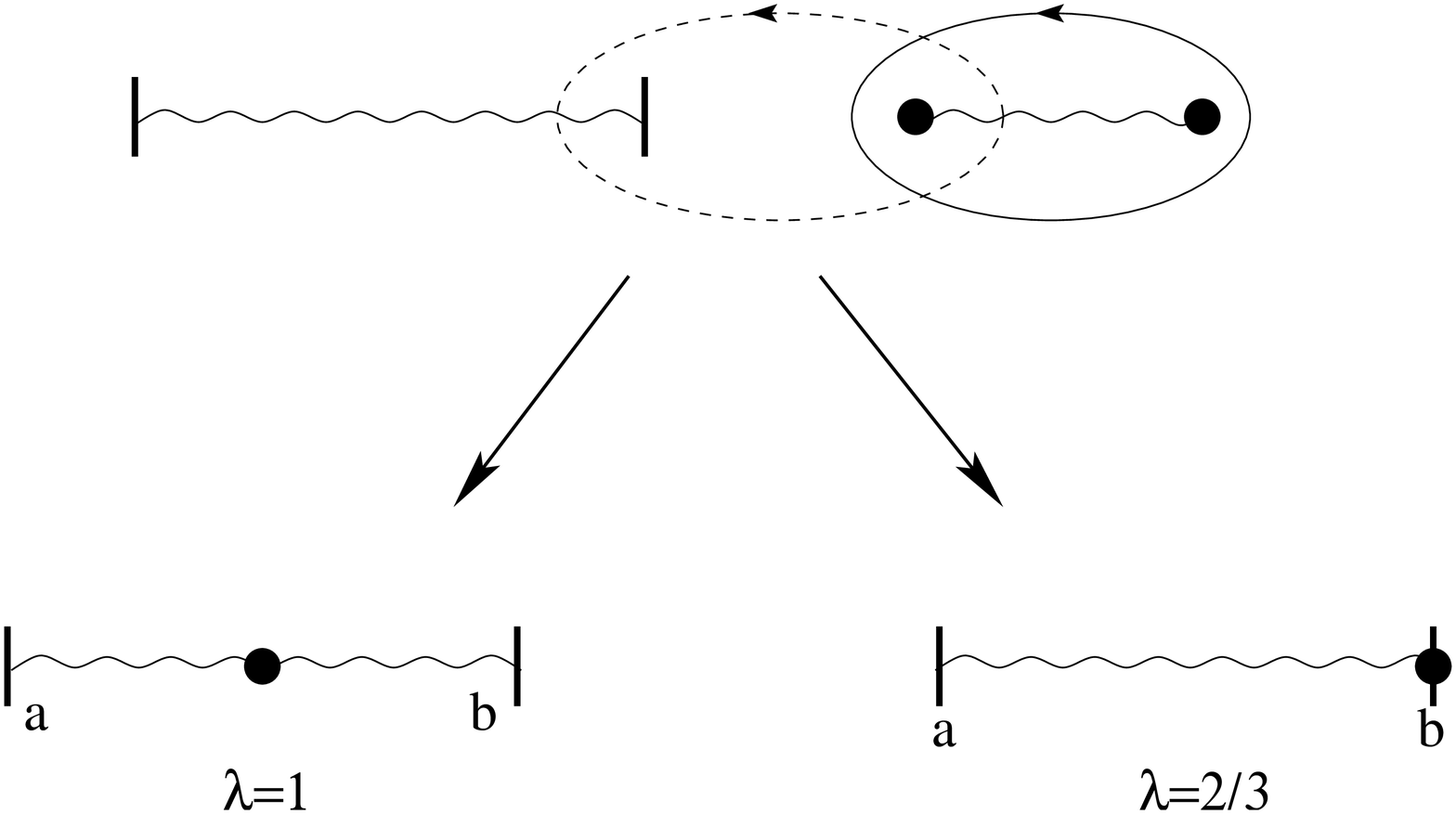,width=8cm,angle=0}%
         }
\newcaption{\footnotesize A schematic illustration of the degeneration
  of two specific cycles within the generic \ntwo U($N$) curve. We can interpret 
  the point $\la=1$ as corresponding to the degeneration of one of the
  cycles of the  generic curve, while the point $\la=2/3$ corresponds
  to the degeneration of two such cycles. In contrast, the 1-cut saddle-point for
  the matrix model is always characterised by one non-degenerate cycle and a
  double point at $x_*$. As shown, at $\la=1$ the double point sits at
  the midpoint of the remaining cut, while for $\la=2/3$ it sits at one end.}}}

The point $\la=2/3$ is, however, known to correspond to
a critical point within the $1/M$-expansion of the matrix model \cite{f2,gzj}, i.e. of the 
prepotential. This suggests that it should be possible
to see consequences of the degeneration
within (\ref{WSmm}).  At $\la=1$, this is of course reflected in the branch
point for ${\cal W}_0(\la)$. Evidence of the degeneration at
$\la=2/3$ is less apparent, but can be uncovered if we define the following
`$\beta$-function' for the period of the curve (\ref{surf}),
\be
 \beta_k \equiv \left. S \frac{\ptl \ta(S)}{\ptl S} \right|_{k},
\ee
where the period (\ref{tak}) is given (off-shell) 
by $\ta(S) = \ptl^2_S {\cal F}$ in terms of the prepotential. Evaluating
$\beta(\la)$ in the relevant $k=0$ intermediate vacuum, we find
\be
 \beta_0 = 2N \frac{1-\la}{3\la-2},
\ee
which vanishes for $\la=1$, and diverges for $\la=2/3$. 
This direct connection between degenerations of the geometry and 
the position of the CMS for large $N$ is rather suggestive, but
without a clearer picture of the physical excitations associated with ${\cal
 W}_{\rm eff}(S)$ it is difficult to fully interpret the consequences
of the latter singularity\footnote{As noted recently \cite{shih},
another quantity with the same characteristics is the formal 
`mass term' ${\cal W}_{\rm eff}''(S)$ which, however, is again
difficult to interpret without knowledge of the K\"ahler potential.},
and we will not pursue this further here.

In the next section, we turn to the dynamical question of deducing the
multiplicity of states, and indeed whether discontinuities do indeed
arise on crossing the CMS curves discussed above.

\section{BPS Wall Spectrum}

The multiplicity of 1/2-BPS multiplets in \ntwo theories in 1+1D
with fixed boundary conditions is given by the CFIV index, $\nu_{jk}$, 
formally defined as \cite{cfiv}
\be
 \nu_{jk} \equiv {\rm Tr}_{jk} F (-1)^F,
\ee 
where $F$ is the fermion number of the corresponding state. The
crucial property of $\nu_{jk}$, following from its definition, 
is its stability under various deformations of the theory. In
particular, it was shown in \cite{cfiv} that it is stable under
(nonsingular) variations of the $D$-terms, such as the K\"ahler
potential, which are unconstrained 
by supersymmetry. This makes calculation of $\nu_{jk}$ tractable as
we can deform the theory, for example via compactification as here,
and the result can be shown to be independent of the compactification
radius $R$ which enters the K\"ahler potential. 

Moreover, $\nu_{jk}$ is also stable under certain variations of the
$F$-terms. For a Wess-Zumino model, characterised by a K\"ahler
potential ${\cal K}(X^a,\bar{X}^a)$ and a superpotential 
${\cal W}(X^a)$, as we have here
this can be understood via a geometric reformulation of $\nu_{jk}$ \cite{cv}.
Specifically, an important consequence of the Bogomol'nyi equation
for 1/2-BPS walls is that the superpotential must trace out a straight
line in the ${\cal W}$-plane \cite{fmvw,at1}; i.e.
\be
 {\cal W}(X^a) = (1-t){\cal W}_0 + t {\cal W}_1, \label{interp}
\ee
where $t\in [0,1]$ is a fiducial variable parametrising the 
transverse coordinate to the wall, and ${\cal W}_0$ and ${\cal W}_1$
are the corresponding vacua between which the wall interpolates.
The set of solutions to the linearised Bogomol'nyi equation about each vacuum 
defines a cycle $\triangle_j$ in field space, and one can show
\cite{cv} that by comparing the two cycles $\triangle_j$ and
$\triangle_k$ at a given value of $t\in [0,1]$, the number of
solutions is given precisely by the intersection number,
\be
 \nu_{jk} = \triangle_j \circ \triangle_k.
\ee
The topological character of the intersection number then implies that 
$\nu_{jk}$ is stable under deformations of ${\cal W}$ which do
not lead to additional vacua crossing the straight line trajectory
between ${\cal W}_0$ and ${\cal W}_1$, as it is only in this
case that the intersection number varies according to a
Picard-Lefschetz monodromy \cite{cv}. 

Thus, to prove the existence of 1/2-BPS states it is sufficient to
verify the presence of smooth profiles in field space for which
the interpolating trajectory in the punctured ${\cal W}$-plane -- with 
additional vacua excised -- is homotopic to the straight line
(\ref{interp}). With this argument in mind, and given our knowledge
of the structure of the CMS curves, our strategy will
be first to determine the spectrum at a convenient point -- we
will consider the pure SYM regime $\la \rightarrow 0$ -- and then
to study whether this multiplicity is preserved on crossing the CMS.
For the latter test, we construct the leading order interaction potential,
as a function of $\la$, between the constituents of the 2-wall
composite discussed in the previous section.

\subsection{Multiplicity for $\la \ll 1$}

We will first consider the spectrum at a special point
$\la=0$, where the infrared theory reduces to pure \none SYM. BPS
walls in this theory have been studied from several points of view 
\cite{walls,walls1,ksy,wallsN,btv,CM}. Within the low energy description
on  $\R^3 \times S^1$, the effective superpotential reduces in 
the $\la \rightarrow 0$ limit to the following
(complexified) affine-Toda form \cite{sw96,kv,hg,ahiss,bho,dorey,dkm,dhk,ritz},
\be 
 {\cal W}_{\rm eff}(\la=0) = \La_{N=1}^3 \sum_{a=1}^N x_a, \label{atlim} 
\ee
subject to the constraint that $\prod x_a=1$. Using the cyclic
symmetry, we choose the initial vacuum to be $x_a=1$ and the second 
vacuum to be $x_a=e^{2\pi ik/N}$, with $k<N$. Within the wall, 
the winding number of each field must then vary from zero to $k/N$ mod $\Z$, and
moreover the constraint $\prod x_a=1$ implies that at any particular
point along the trajectory (\ref{interp}) the
winding numbers of the fields $x_a$ must sum to zero. Thus, at any
non-vacuum point, they cannot all be equal. To minimise gradient
energy, there are only two allowed winding
numbers: $k/N$ and $k/N-1$ \cite{rsv}, and this
leads us to the following resolution of the constraint in terms of
a new variable $y$ \cite{hiv},
\be
 x_{a=1,\ldots,k} = y^{N-k}, \;\;\;\;\;\; x_{a=k+1,\ldots,N} = y^{-k}.
\ee
The relation (\ref{interp}) then implies that for the corresponding $k$-walls
\be
   k y^{N-k} + (N-k) y^{-k} = N(1-t) + N t e^{2\pi i k/N}.
\ee
For $N=2$ (and thus $k=1$), this relation can be straightforwardly 
inverted to obtain the interpolating trajectories. However, for
generic $N$, we can instead simply verify the existence of BPS solutions
by following an approach used in the context of the \ntwo affine-Toda 
model in 1+1D\footnote{As an aside, bosonic affine-Toda theory in 1+1D with an 
imaginary coupling is also known to exhibit solitons \cite{holl}. In general,
these solutions satisfy a second-order equation with the potential
(\ref{atlim}), and so are not related to the configurations we are
concerned with here. However, the case $N=2$ is an exception where the
sine-Gordon soliton is (up to certain rescalings) a BPS solution if we
use the classical K\"ahler metric. One may understand this via noting 
that affine-Toda solitons also satisfy a set of first-order
(B\"acklund) equations \cite{lot} which reduce precisely to the 
Bogomol'nyi equations for $N=2$.} \cite{hiv}. Note that since
$y$ is a complex variable, there are two linearised solutions
about each vacuum, but at most one combination of these can link to form a 
BPS trajectory. To verify that such a trajectory exists we follow
the argument described above \cite{cv,hiv}. In particular, with
the ansatz $y(t) = e^{2\pi i t/N}$, we find
\be
 |{\cal W}_{\rm eff}(\la=0)| = |N-k + ke^{2\pi i t} | \leq N,
\ee 
where the final equality holds only at the initial and final vacua.
Therefore, since the remaining vacua lie on a circle of radius $N$, this 
suffices to prove the existence of the corresponding BPS 
multiplet \cite{hiv}. Moreover, accounting for the permutation 
symmetry in the identification of $y$, the full multiplicity is
given by 
\be
 \nu_{0k}(\la=0) = \left( \begin{array}{c} N \\ k \end{array}\right).
 \label{mult}
\ee
This result, conveniently interpreted as a multiplet transforming in
the $k^{th}$ fundamental representation of SU($N$), 
is consistent with other calculations of the degeneracy of
BPS walls in pure \none SYM \cite{av,rsv}.

\subsection{Inter-wall potential for generic $\la$}

To explore how the multiplicity of BPS states varies as we move in 
parameter space, it is useful to follow a somewhat different approach.
It should be clear from the analysis above that the spectrum
of 1-walls will be stable for $|\la|<1$ where the structure of
neighbouring vacua does not change qualitatively. However, it is less
apparent that the multiplicity of higher $k$-walls is stable in this
region of the parameter space. We can determine possible
discontinuities by treating $k$-walls as bound states of $k$ 1-walls
and calculating the inter-wall potential as a function of $\la$.
In practice, it will be sufficient to focus on the simplest case of 
a 2-wall bound state, corresponding to the example considered in 
the previous section.

To proceed, we note that on general grounds the only
global symmetries broken by the wall configurations are
super-translations. We then infer that the only bosonic
moduli of a system of two asymptotically separated 1-walls will be
their respective centre-of-mass positions. It is an implicit
assumption that any other worldvolume
fields are massive and can be ignored for the purpose of considering
the ground state. The bosonic moduli space of 
the constituent system is then $\R^2$, and we can always decouple 
the overall translational mode, and thus the relative moduli 
space is one-dimensional, parametrised by a real field $r$. 
The problem at hand then reduces to computing the potential
induced on this space when the constituent 1-walls are at a finite 
separation.

For the region $|\la|<1$, it convenient to further restrict the 
problem by considering the regime where 
$R\La_{N=2} \ll 1$
which means that the adjoint scalar fields $p_a$ are (generically) of
higher mass than the gauge modes $q_a$ and 
can be integrated out, leading to the reduced superpotential,
\be
 {\cal W}_{\rm eff}(x) = \frac{2N}{3\la}\La_{N=1}^3 \left( 1 - \frac{1}{N}
 \sum_{a=1} \left(1-\frac{\la}{2}\left( x_{a} + x_{a-1}\right)
     \right)^{3/2}\right),
\ee
which is now a function only of $x_a$. However, the presence of branch
points implies that this reduced system is valid only for
sufficiently small $\la$, for which the field profiles remain well away
from the cuts.

As noted above, we focus on 2-walls and fix the two vacua
to be $x_a^{(0)}=e^{-2\pi i/N}$ and $x_a^{(1)}=e^{2\pi i/N}$ as in Section~3. The
putative 2-walls in this sector, counted by $\nu_{-11}$ can be viewed 
as bound states formed from the 1-walls for which $\nu_{-10}=N$ and
$\nu_{01}=N$. A schematic illustration of how the phases of the
$x_a$ fields must vary over the wall profile is shown in Fig.~5.

\FIGURE[t]
  {\parbox{10cm}{
    \centerline{%
   \psfig{file=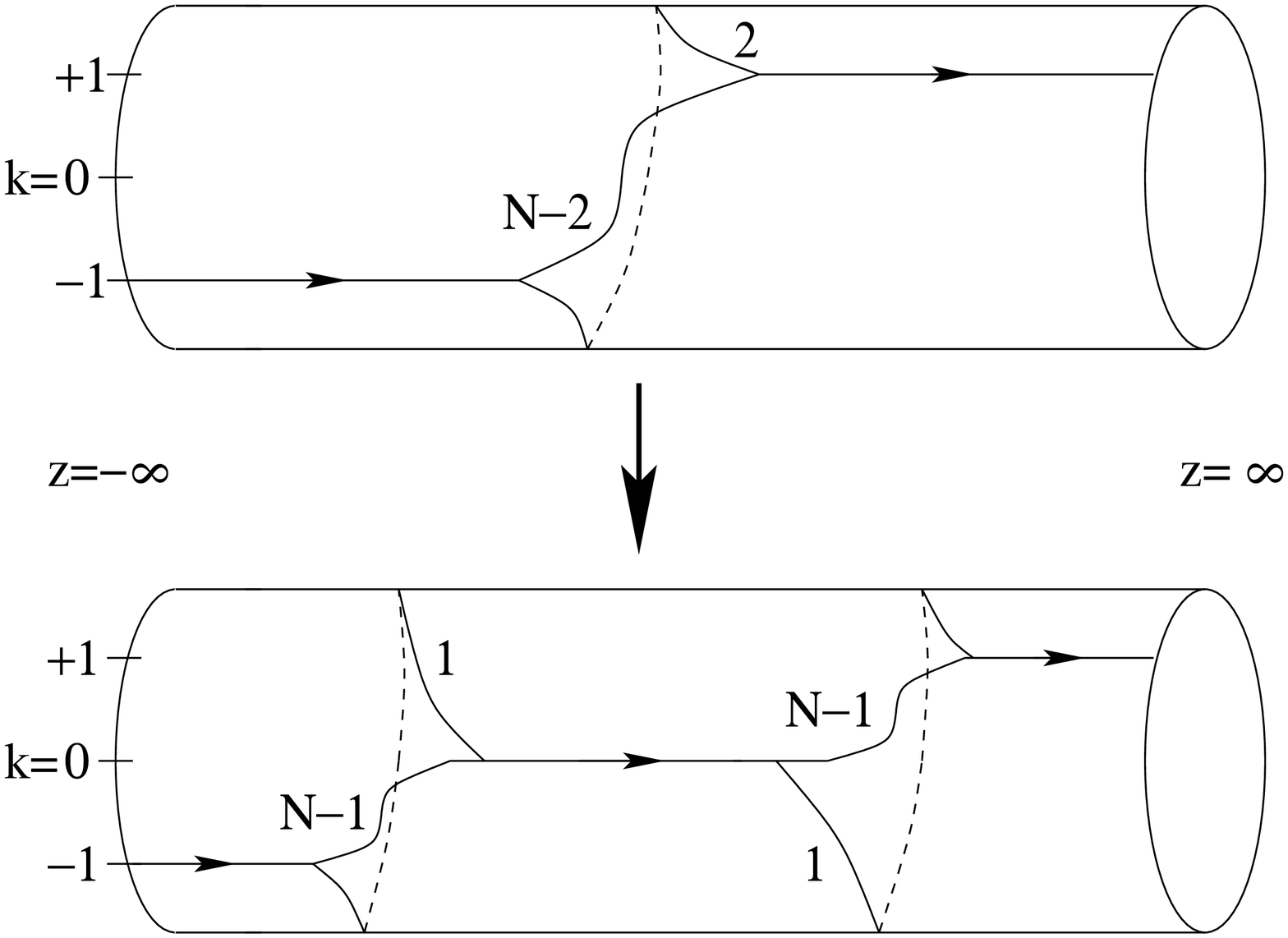,width=10cm,angle=0}%
         }
\newcaption{\footnotesize A schematic
representation of the profile of the phase of the $x_a$ fields within
 the 2-wall bound state in the upper plot. $N-2$ of the components have
 winding number $2/N$, while the remaining two have winding number
 $2/N-1$. In the lower plot the two 1-wall constituents are separated
 so that the fields all lie near the intermediate vacuum in the
centre of the 2-wall. This configuration is unstable but can be used
to extract the leading order interaction potential.}}}

The leading order potential between the constituents can be
determined by expanding the tension of wall two in the background
of wall one (see also \cite{rsvv,pt}). 
More precisely, we can write the tension of the
second wall $T_2 = 2 |{\cal W}_{(1)} - {\cal W}_{(0)}|$ as
\be
 T_2(r) = T_2 - 2{\rm Re}\left[
e^{-i\ga}\de_{(-10)}^a(z_0)\de_{(01)}^b(z_0+r)\ptl_a\ptl_b {\cal
W}_{(0)}\right] +\cdots,  \label{t2r}
\ee
where the second term is the leading order interaction potential
$V(r)$ between the constituent 1-walls positioned at $z=z_0$ and
$z=z_0+r$ respectively (see Fig.~5), with $z$ the transverse coordinate.
In this expression, the subscripts label the phase of $x_a$ in
the relevant vacua, and $\de^a = x^a - x^a|_{(0)}$ are the deviations
of each field from its value in the intermediate vacuum, while $\ga$
is the phase of ${\cal Z}_{01}$. These
fields satisfy the linearised Bogomol'nyi equations given by
\be
 \ptl_z \de^a = e^{i\ga} M_b^a \bar{\de}^b, \;\;\;\;\;\;\;{\rm where}\;\;
 \sum_{a=1}^N \de^a = 0, \label{cons}
\ee
and $M_b^a = g^{ac} \ptl_c\ptl_b {\cal W}_{(0)}$ is the (complex)
mass matrix in the intermediate vacuum, with $g_{ab}$ 
the K\"ahler metric. The additional constraint
on the perturbations follows from $\prod x_a=1$. 

The Bogomol'nyi equations decouple in the mass-eigenstate basis for
the intermediate vacuum, where $M$ is diagonal. In other words,
since $M$ is symmetric, the eigenvalues of the following system \cite{cv}
\be
 \left( \begin{array}{cc} 0 & e^{-i\ga}\bar{M} \\ e^{i\ga}M & 0
 \end{array}\right) \left( \begin{array}{c} \de \\ \bar{\de}
 \end{array}\right) = \pm m_{\mu} \left( \begin{array}{c} \de \\ \bar{\de}
 \end{array}\right),
\ee
are real and paired $\{m_\mu,-m_{\mu}\}$, for $\mu=1,\ldots,N-1$, and moreover
$\{m_\mu\}$ can be identified with the mass eigenvalues in 
the intermediate vacuum. We deduce that the
complex phase of the perturbations $\de$ arises purely from the integration
constant, and can be fixed via comparison with (\ref{cons}). 

With explicit solutions for the linear perturbations in hand, it is convenient
to first use the Bogomol'nyi equation to write the potential in the
form
\be
 V(r) = - 2 {\rm Re}\left[ \de_{(-10)}(z_0) \cdot
 \bar{v}_{(01)}(z_0+r)\right] + \cdots,
 \;\;\;\;\;\;\;\;{\rm where}\;\; v^a = \ptl_z \de^a,
\ee
and the inner product is that associated with the K\"ahler metric.
On substituting the appropriate solutions, we can read off 
the leading order potential,
\be
 V(r,\la) = - 2 \sin\left(\frac{\om(\la)}{2}\right) \sum_{\mu=1}^{N-1} m_\mu
e^{-m_\mu r}  +\cdots, 
\ee
where $\om(\la)$ is the relative phase between the central charges of
the two constituent 1-walls, as defined in (\ref{omdef}).

Note that the constraint apparent in the fundamental weight basis
in (\ref{cons}) is reflected in the mass spectrum. For example,
in the limit $\la=0$, we find
\be
 M^a_b(\la=0) \propto \hat{C}^a_b  \;\;\;\;
 \Longrightarrow \;\;\;\; m_{\mu}(\la=0) \propto \sin^2 \frac{\pi \mu}{N},
\ee
where $\hat{C}^a_b$ is the Cartan matrix for affine SU($N$). The
low-lying mass eigenvalues then appear in pairs, and depend
nontrivially on $N$. 

It is now clear that the presence of the corresponding 2-walls depends
purely on the value of $\om(\la)$, which is kinematic in nature
as one should expect, since discontinuities in the BPS spectrum are
allowed only when the corresponding bound states are marginally
stable, namely when $\om(\la)=0$. Near these submanifolds the potential reduces to
\be
 V(r,\la) = - \om(\la) \sum_\mu m_\mu e^{-m_\mu r} +\cdots,\label{Vr}
\ee
which is linear in $\om$ and thus a discontinuity in the spectrum
should apparently occur on crossing the curve. Moreover, the potential
applies to all components of the multiplet (\ref{mult}), and so it
would seem that the entire multiplet is removed from the
spectrum when $\om(\la)<0$. However, before reaching this conclusion, we first need
to consider the effect of quantum corrections.

\subsection{Quantum corrections}

Before drawing conclusions regarding discontinuities in the spectrum,
it is important to understand the status of the leading order
potential that we have derived. In particular, we are interested in
submanifolds in parameter space on which the bound state is marginally
stable, where $\om(\la)=0$, and the arguments above suggest that
we can equivalently define the submanifold via the relation
$V(r,\la)=0$. However, it turns out that the latter relation is not stable
to quantum corrections, and more work is required. In actual fact 
quantum corrections, associated with the fermionic content of the 
model, are very important and become increasingly so near the CMS \cite{rsvv}. 
Fortunately, one finds that the qualitative insight drawn from 
$V(r,\la)$ is nonetheless correct \cite{rsvv}. Indeed, this must be the case 
on the grounds that the condition $\om(\la)=0$ describing the CMS is kinematic and 
protected by supersymmetry.

To proceed, we will compactify the spatial worldvolume
coordinates on an additional circle of sufficiently small radius $L$. The low energy
description of the worldvolume theory, i.e. for scales well below
$1/L$, then reduces to quantum mechanics on a one-dimensional moduli
space coordinatised by $r$. In this regime, the leading order potential (\ref{Vr}) is 
corrected by the fact that walls are 1/2-BPS, and
thus the (dimensionally reduced) worldvolume theory must 
possess \ntwo worldline supersymmetry. 
Such quantum mechanical systems are well understood \cite{EWi}.
If we denote by $\{Q_1,Q_2\}$ the supercharges which act trivially
on the putative BPS bound state, i.e. 
\be
 (Q_1)^2 = (Q_2)^2 = (RL) (T - |Z|) \ll (RL) T, \label{rel}
\ee
we expect to find a bound state if this system exhibits a
unique vacuum; the multiplet structure is then reproduced
on tensoring this state with the free centre-of-mass sector.

The generic structure of \ntwo SQM was first
described by Witten \cite{EWi}, and we can realise the algebra 
as follows in terms
of a quantum mechanical superpotential ${\cal W}_{\rm QM}(r)$,
\bea
 Q_1 &=& \frac{1}{\sqrt{2M_r}}\left[\pi_r\si_1 
  + {\cal W}'_{\rm QM}(r) \si_2\right], 
 \nonumber\\
 Q_2 &=& \frac{1}{\sqrt{2M_r}}\left[\pi_r\si_2 - 
  {\cal W}'_{\rm QM}(r) \si_1\right], \label{scharges}
\eea
where $M_r=(RL)(1/T_1+1/T_2)^{-1}$ is the effective reduced mass.
The Hamiltonian is then given by 
\be
 {\cal H}_{\rm QM} = M - |Z| = \frac{1}{2M_r} \left[\pi_r^2 
+({\cal W}'_{\rm QM}(r))^2 + \si_3 {\cal W}''_{\rm QM}(r)\right], \label{Hsqm}
\ee
where the second term is the classical potential, and the
final term is a quantum correction of ${\cal O}(\hbar)$. 

We have not explicitly allowed any nontrivial corrections to the kinetic term.
In general one would expect that massive exchanges will introduce
such terms, i.e. the coefficient of $\pi_r^2$ should take the schematic
form $g_{rr} = 1 + {\cal O}(e^{-mr})$. However, by a rescaling, $g_{rr}$
can always be absorbed into the potential, and this is implicitly
the basis used here since we will in practice match to the full bosonic potential
$V(r) = g^{rr} ({\cal W}'_{\rm QM})^2$.

Identifying the quantum mechanical superpotential ${\cal W}_{\rm
QM}(r)$ in general requires a detailed analysis of the model
at hand, as in \cite{rsvv}. However, we can easily extract the 
general structure near the CMS by comparison with the leading order classical
potential in (\ref{Vr}) \cite{r2}. Having compactified the theory on $T^2$, we can
augment (\ref{Vr}) with the leading constant term corresponding to
the binding energy of the two constituents. Near the CMS, the
bosonic potential then has the generic form,
\be
 V(r,\la) = \frac{1}{2}M_r \om(\la)^2 - \om(\la) \sum_\mu m_\mu
 e^{-m_\mu r} +\cdots. \label{Vr2}
\ee
Comparing (\ref{Hsqm}) with (\ref{Vr2}), we find
\be
 {\cal W}_{\rm QM}(r,\la) = \om(\la) (M_r r) 
   + \sum_\mu \exp(-m_\mu r) + \cdots, \label{Wsqm}
\ee
and thus near the CMS,
\be
 V_{\rm QM}(r,\la) = \frac{1}{2}M_r \om(\la)^2 
 - \sum_\mu m_\mu \left[\om(\la) +\si_3 \frac{m_\mu}{2M_r}\right] \exp(-m_\mu r) 
 + \cdots
\ee
The additional quantum correction to the leading term becomes
important near the CMS where it remains finite.

Despite this correction, the classical equilibrium position survives as
the maximum of the ground state wavefunction, which is easily
determined from (\ref{scharges}),
\be
 |\Ps_0\rangle = \exp(-{\cal W}_{\rm QM}(r)) | - \rangle  
 \stackrel{r \rightarrow \infty}{\longrightarrow}  \exp(-\om(\la) M_r
r) | - \rangle +\cdots 
\ee
where $|\pm \rangle$ are the eigenvectors of $\si_3$.
In the limit $r \rightarrow \infty$ we have extracted only 
the leading exponential behaviour. It is apparent that 
this wavefunction is normalisable on only one side of the CMS, namely
for $\om>0$, which determines the existence domain for the BPS bound
state. The intuition regarding discontinuities in the spectrum
drawn from (\ref{Vr}) therefore survives at the quantum level,
although the details are somewhat different \cite{rsvv,r2}.

\subsection{Discontinuities in the BPS spectrum}

The analysis of the leading order potential allows us to reduce 
questions about the 2-wall multiplicity to purely kinematic issues 
concerning the dependence of the central charges on $\la$. In
particular, as discussed in section~3, discontinuities only occur 
on co-dimension one submanifolds of the parameter space -- 
curves of marginal stability -- where $\om(\la)=0$. The analysis of
this section has verified that such discontinuities do indeed 
occur in this case. 

Following our discussion in section~3, it is convenient for
illustrative purposes to take $N$ large, so as to resolve the 
branch structure of the central charges as functions over the
$\la$-plane. To leading order in $1/N$, we recall that the angle
$\om$ takes the simple form
\be 
 \om(\la) = \frac{2\pi}{N} \left[ 1- \frac{1}{2} {\rm Re}
  \left( \frac{\la}{1-\la}\right)\right] 
  + {\cal O}\left(\frac{1}{N^{2}}\right),
\ee
and one readily verifies that $\om$ is positive, and thus the
BPS 2-wall bound states indeed exist outside the closed circle defined
by $|6\la - 5|^2 =1$, while the potential becomes
repulsive, and the BPS bound states disappear from the spectrum 
in the interior shaded region of Fig.~6.

\FIGURE[t]
  {\parbox{10cm}{
    \centerline{%
   \psfig{file=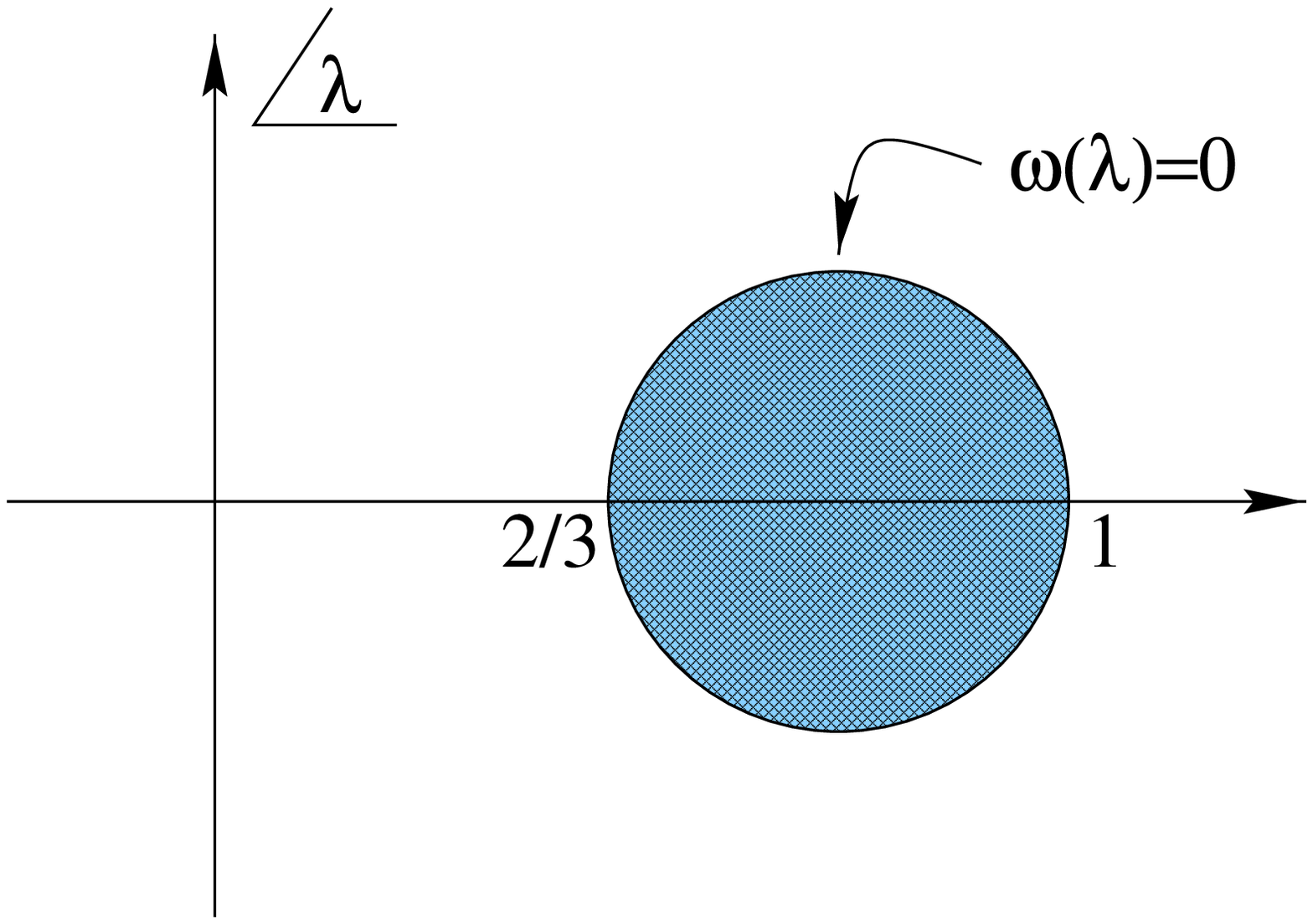,width=6cm,angle=0}%
         }
\newcaption{\footnotesize A plot of the 
  circular (large $N$) CMS curve in the $\la$-plane for the 
  2-wall composite. The 2-walls are absent in the shaded interior region.}}}

In concluding this section, we will attempt to draw one further conclusion
from these arguments relevant to the $\la\sim 1$ regime, although
this is strictly outside the regime of validity of the dynamical
approach being used. Indeed,  it is apparent that the expression for the 
relative phase at large $N$
is singular at $\la=1$. This is the critical point at which the 
intermediate vacuum ${\cal W}_0$ collides with the corresponding
confining vacuum ${\cal W}_N$ on
the second branch, signalling the presence of additional light
states. Moreover, for $N$-even, $\la=1$ is also a Seiberg-Witten singularity
and denotes a point where the confining branch intersects the
Coulomb branch.

For the composite 2-walls considered above, this suggests the
possibility of forming a sandwich-like configuration where the outer
regions are in confining phases -- each associated in the near
\ntwo regime with condensation of charge-one dyons and anti-dyons
respectively -- while the interior domain, between the two constituent
1-walls, is in a massless
phase at $\la=1$. Such configurations would be of interest as there are then
generic arguments for the formation of flux-tubes \cite{ds3}, i.e. open strings
ending on the wall, and other features reminiscent of $D$-branes (see
e.g. \cite{sy}). However, although not strictly valid for $\la=1$, the 
arguments above suggest that, at least within this system, there is no such BPS configuration.
i.e. as we tune $\la \rightarrow 1$ along a trajectory outside, but
close to, the CMS the composite delocalises
and on reaching $\la=1$ only the massless phase remains.

\section{Discussion}

We have presented a limited exploration of the spectrum of BPS domain
walls within \none gauge theories with adjoint matter, and
specifically those characterised by a space of
parameters associated with a classical polynomial superpotential. We have been
far from exhaustive and focused on a small subset of states which in 
an appropriate limit reduce to the BPS walls present in \none SYM. The
aim was simply to exhibit some of the features of the spectrum which,
through the interpretation of the wall multiplicity in terms of the
CFIV index, provides `chiral data' on the parameter space. 

In this final section, we will return to some of the original
motivations noted in Section~1, and comment on some extensions.

\subsection{Embedding within \nones}

One motivation for studying the BPS spectrum was to gain insight into
additional symmetries (or dualities) acting on the parameter
space. With the aim of gaining a clearer picture of the quantum symmetries
preserved by the BPS spectrum, it is helpful to consider the natural
UV completion of the theory studied here within \nfour SYM, a theory which exhibits
$S$-duality. Perturbing \nfour SYM via the addition of a cubic 
superpotential for the three adjoint chiral fields formally preserves modular
invariance \cite{intril,adk} if we assign the mass terms 
modular weight $(-5/6,1/6)$, and the cubic couplings weight $(-1,0)$. 
One may then construct the low energy effective superpotential by
compactifying on $\R^3 \times S^1$ \cite{dorey}, and using the
correspondence with the elliptic Calogero-Moser integrable system
\cite{martinec,dw}.
In particular, the gauge invariant monomials are again identified
with the action variables, 
$\left< {\rm tr}(\Ph^p)\right> \longleftrightarrow {\rm tr}(L^p)$,
where $L$ is the corresponding Lax matrix \cite{op}. 

Suppressing the details, we can follow a similar procedure to that presented in
Section~2 to determine the massive vacua. Restricting to those which
correspond to a classically unbroken U($N$) gauge group, the central
charges have the form \cite{mstar},
\be 
 {\cal Z}_{ij} = 2 \sum_k n^{(ij)}_k {\cal W}_k
 = \frac{4N}{3\la} \La_{N=1}^3 \left[ (1-\xi X_j)^{3/2}
- (1-\xi X_i)^{3/2}\right],
\ee
where 
\be
 \xi = 8 g^2 \frac{M^2}{m^2}, \label{xidef}
\ee
in terms of the mass parameters for the three chiral fields $m_i=(m,M,M)$, with
$m \ll M$, and
\be
 X_k = \frac{1}{24}
\left[C(\ta)-\frac{p}{q}E_2\left(\frac{p\ta+k}{q}\right)\right], \label{Xk}
\ee
where $(p,q)=(1,N)$, with $k=0,\ldots,2N-1$ for the confining vacua,
and $(p,q)=(N,1)$, with $k=0,N$ for the Higgs vacuum. 
$E_2$ is the 2nd Eisenstein series, and $C(\ta)$ is a vacuum-independent constant.
With an appropriate choice of $C(\ta)$, which corresponds to a
particular basis for the action variables of the Calogero-Moser
system, the combination
$\xi X_k$ is modular invariant up to permutations of the vacua, and
so the central charges have weight $(-1/2,1/2)$. Consequently, the
wall tensions are invariant up to permutation, and SL(2,$\Z$) thus acts via
permutations on the wall spectrum.

The parameter space in this theory is now two-dimensional,
coordinatised (for the massive U($N$) vacua) by $\{\xi,\ta\}$. One can
show that in the decoupling limit, $M\rightarrow \infty$, 
the spectrum of 2-walls interpolating between
confining vacua exhibits the same CMS curve as found above, but whose 
profile is (not surprisingly) corrected by fractional instanton
effects. The permutation symmetry of the vacua under rotations of 
$\la$ is now seen to follow directly from the symmetry under 
$T$-translations, namely
$\ta\rightarrow \ta+1$, in \slz. A more complete analysis of the spectrum
and symmetries, including the `classical' walls connecting confining
and Higgs vacua, will appear separately \cite{mstar}.

\subsection{On the anomaly multiplet}

Another motivation for this work was to try and understand whether
the matrix model approach could be used to extract the chiral data
associated with the BPS spectrum -- namely to compute the CFIV index --
in a purely four-dimensional context. An important
point to recognise is that, while the confining vacua are described by 
1-cut solutions to the matrix model, this is insufficient to describe 
BPS states. Indeed, the field profiles required to construct 
domain walls will necessarily pass through generic regions in moduli
space where, in the \ntwo limit, the gauge group is maximally broken.
This implies that we require knowledge of the large $M$ solution to
the matrix model with the maximal number of cuts. On reflection, this
should not be too surprising on comparison with the approach used 
here, where we made use of the full \ntwo Seiberg-Witten curve.

One may then enquire as to how the nontrivial wall multiplicity, described in
section~4 \cite{av,rsv}, arises in the pure \none SYM limit
where, having integrated out the adjoint field, the effective superpotential
${\cal W}={\cal W}(S)$ depends only on a single gluino
condensate field $S$.  Even discounting non-analyticities, 
a superpotential depending on a single field can at most describe
unique interpolating solutions, and this is insufficient to explain
the multiplicities found in section~4. One may then ask how the
information about this multiplicity is encoded in four dimensions,
if at all? To this end, it is worth recalling the resolution of a (perhaps) analogous
puzzle that arises on compactification to 1+1D on a 2-torus (see also
\cite{aivw}). Since we are computing the index ${\rm tr}\,F(-1)^F$, one might
anticipate that such an additional reduction should not 
affect the conclusions. 

We can ignore the presence of the adjoint chiral field,
so the reduced superpotential follows directly from a reduction of the
3D affine-Toda superpotential,
\be 
 {\cal W}_{\rm 2D}(\Si,v_a) = \La_{2D} \sum_{a=1}^N e^{-v_a} 
 + \Si \left(\sum_a v_a\right),
 \label{W2d}
\ee 
where $x_a=e^{-v_a}$, with $\La_{2D}=4\pi^2(RL)\La_{N=1}^3$, 
and we have inserted a Lagrange multiplier to explicitly enforce
the decoupling of the overall U(1) factor. One can recognise this
as the 1+1D mirror \cite{hv} of the \ntwo $\C P^{N-1}$ sigma model where we
interpret $\Si$ as the U(1) field strength within the corresponding
linear sigma model. This relationship is also consistent with a
direct reduction of pure SYM, since the moduli space of flat SU($N$) 
connections on a torus is also $\C P^{N-1}$ \cite{redux}. 

If we now integrate out $\Si$ from (\ref{W2d}), we simply enforce the
decoupling of the central U(1), and we can reproduce the vacuum
structure and nontrivial kink multiplicity in the same manner as
described in section~4 \cite{hiv}. However, we could also choose
to integrate out the $v_a$'s, or equivalently, the homogeneous
$\C P^{N-1}$ chiral fields. From (\ref{W2d}), we can do this at
tree level to find,
\be
 {\cal W}_{\rm 2D}(\Si) = \Si \left( \ln \frac{\La_{2D}^N}{ \Si^N}
 +N\right). \label{cpn}
\ee
This reduced theory describes, as it must, the same vacuum structure
but, due to Gauss law, actually cannot describe any nontrivial kinks
as one must satisfy the constraint $\Si |_{-\infty} = \Si |_{+\infty}$ 
\cite{witten}. As noted by Witten, the resolution of this puzzle is that the kinks 
are actually charged under the additional $\C P^{N-1}$ fields that have been 
integrated out, and one must account for the coupling of $\Si$ to the corresponding current
\cite{witten,hh}. In this sense, the superpotential (\ref{cpn}), while
sufficient for describing the vacua, is insufficient to fully describe
the BPS spectrum.

Although there is no direct analogue of the Gauss law constraint in
3+1D, where we lift (\ref{cpn}) to the Veneziano-Yankielowicz 
superpotential \cite{vy}, one could make quite a close analogy if we 
were to interpret $S$ as the field strength of a linear multiplet \cite{linear}
containing a 3-form. While we will not pursue this further here, 
this issue is intriguing as these kinks are precisely the dimensional
reduction of the BPS wall configurations that we have been counting.

\acknowledgments

I would like to thank Jan de Boer, Mikhail Shifman, and Arkady Vainshtein for
helpful discussions and/or comments on the manuscript, and especially
Nick Dorey for discussions and correspondence, some
years ago, on BPS walls in compactified gauge theories.


\end{document}